\documentstyle[12pt]{article}
\input epsf
\newcommand{\svec}[1]{\mbox{\boldmath $#1$}}
\begin{document}
\title{ 
\tt Raynal's use of the word ``aberrant'' 
appears more appropriate for
his ECIS formulation}
\author{K. Amos$^{1}$, L. Canton$^{2}$, G. Pisent$^{2}$, 
J.P. Svenne$^{3}$, D. van der Knijff$^{4}$}
\date{  }
\maketitle
\centerline{\small \it $^1$ School of Physics, University of Melbourne, 
Victoria 3010, Australia}
\vskip 0.2cm
\centerline{\small \it $^2$ Istituto Nazionale di Fisica Nucleare,
Sez. di Padova e} 
\centerline{\small \it Dipartimento di Fisica dell'Universit\`a, 
Padova, 35131 Italy}
\vskip 0.2cm
\centerline{\small \it $^3$ 
Department of Physics and Astronomy, University of Manitoba, and}
\centerline{\small \it Winnipeg Institute for Theoretical Physics, 
Winnipeg, MB R3T 2N2, Canada}
\vskip 0.2cm
\centerline{\small \it $^4$ Advanced Research Computing, Information Division,
University of Melbourne, Victoria 3010, Australia}
\begin{abstract}
Recently, we published a paper (Nucl. Phys. A 728 (2003) 65)
presenting a new calculational method for 
nucleon-nucleus elastic scattering at low energies. That method
is particularly appropriate for analyses in the region of 
narrow resonances.
The method is based upon the sturmian representation of the S-matrix, 
and allows inclusion of nonlocality effects due to Pauli principle.
It also provides a systematic identification of 
narrow-resonance spectra and subthreshold bound states.
A phenomenological test calculation  for low-energy (below 4 MeV) 
neutrons on $^{12}C$ (including the first two excitations of the target)  
was presented to illustrate the validity of the approach.
The model calculation received a violent criticism (see {\tt nucl-th/0312038}) 
by the developer of a method (ECIS) which to date cannot handle 
nonlocality effects and cannot be used easily to identify all
narrow resonances. 
We demonstrate that Raynal's opposition to our development
is not well founded by the arguments he presents. 
Indeed the work we published
shows, on rewording the title of {\tt nucl-th/0312038}, that it 
is ``aberrant'' phenomenologically to 
analyze resonant low-energy nucleon-nucleus processes with 
coupled-channel methods 
without taking into account the nonlocalities due to the Pauli principle; 
problems typically encountered in the ECIS formulation.
\vskip .5truecm
PACS numbers 24.10.-i, 25.40.Dn, 25.40.Ny, 28.20.Cz
\end{abstract}

In the comment of Ref.~\cite{Ray},
it is implied that we generate our 
coupled-channel spin-orbit potential
by taking the standard elastic spin-orbit potential in channel c, 
$({\bf l \cdot s})_c$ and make it ``coupled-channel'' with the phenomenological
prescription:
\begin{equation}
W_{ls} [\ell .\svec {s}]_{c} \delta_{cc'} \rightarrow
{W_{ls} \over 2} \{[\ell .\svec {s}]_{c'}
+[\ell .\svec {s}]_c\}\, . \label{E1}
\end{equation}
This is not the case, but one could be misled so 
by Eq.(1) of Ref.~\cite{Ray}, where one small piece
of a complex formula is taken out of context.

Then Raynal makes an illegitimate comparison of this piece alone,
set apart from the context, with the six-parameter model 
coupled-channel spin-orbit potential used in his code ECIS
and derived in some extremely hard-to-find publications. In his
article, Ref.~\cite{Ray}, he persists 
in the erroneous claim that we do such limited kind of 
coupled-channel generalisation of the elastic spin-orbit term 
[see Eqs.(5-6)].

Given that the false criticism could be disguised by his presentation
we restate here how we generate the coupled-channel ${\bf L\cdot S}$  term
in our phenomenological test calculation and what are the considerations
involved.

The use of a deformed optical potential to reproduce a low energy 
spectrum of resonances is quite a peculiar problem. The literature on
the argument can be traced back  (to our knowledge) 
to  few references \cite{res}.
The problem has been also summarised by P.E.Hodgson in his book
\cite{hod}.
With respect to usual (central or 
deformed) optical model analyses, this specific problem is 
characterised by the low energies involved and the 
discrete structure of the resonance spectrum.
Since the potential parameters collectively influence the resonance spectrum,
the inverse-problem solution (derivation of the potential parameters 
from fitting the background and resonances)  therefore is particularly 
cumbersome.

Ref.~\cite{amo} discusses a particular technique of algebrization of the
multichannel scattering problem through finite--rank 
expansion of the interaction.
The technique is particularly suitable for the problem because it
allows one: (a) systematically to find the narrow and super-narrow 
(compound) resonances through the study of sturmian trajectories,
by-passing the major computing problem of using an extremely fine energy grid, 
and (b) to eliminate the Pauli--forbidden states from the channel couplings.

The last point is particularly important. In table I of Ref.~\cite{amo} 
we compare 
the spectra obtained with and without elimination of the Pauli--forbidden 
states. Pauli blocking matters significantly, and points to a known but
unfortunately frequently forgotten issue: 
when nucleon--nucleus processes are described by
collective degrees of freedom, the associated calculations typically 
disregard the 
antisymmetrization effects. Note that these effects are also relevant for 
DWBA calculations of scattering~\cite{Review}. 

The particular choice made for
the input-potential form in no way detracts from these major points of our 
paper. However, Raynal has misrepresented what we have used and so 
we review the model potential we have employed in our test calculation.
The choice of the phenomenological potential is an old form 
\cite{res}, namely a central part plus a spin--orbit term (with surface--type
radial dependencies) plus spin--spin and orbit--orbit terms (with volume--type
radial dependence). All coefficients are considered as adjustable parameters,
to be determined through the fitting procedure. Raynal objects to the 
spin-orbit aspect arguing for a form derived from reduction of the 
Thomas term. But all are merely phenomenological representations.

Using the notation given in our paper \cite{amo},
we write the Thomas term as follows:
\begin{equation}
V_{th}=-i V_{ls} {\bf S} \cdot (\nabla f \times \nabla),
\label{base}
\end{equation}
where ${\bf S}$ is the nucleon spin operator and $f$ is defined by Eq.~(37) 
of Ref.~\cite{amo} in the case of central interaction.

One can write the Laplacian in the form
\begin{equation}
{\bf \nabla}={{\bf r} \over r} \nabla_r+{1 \over r}{\bf \nabla}_\Omega \ ; \
{\bf L}=-i {\bf r \times \nabla}.
\label{def}
\end{equation}
Then, for central potentials [f=f(r)], the usual expression is easily found,
namely:
\begin{equation}
V_{th}=V_{ls}{1 \over r} {df \over dr} {\bf S}\cdot {\bf L} \, .
\label{ls}
\end{equation}
We now introduce distortion as in the Tamura model~\cite{Tam}. The new
distorted potential is described by equations (38) and (39)
of Ref.~\cite{amo}. If we consider for the moment first-order expansion
in the deformation parameter $\beta$ of the factor $f$ as given in Eq.(12) 
of Ref.~\cite{amo},
we obtain
\begin{equation}
V=V_{ls}\{{1 \over r}  [{df \over dr} +\epsilon
 {d \over dr} {df \over d\epsilon}] {\bf S \cdot L}
 -
{i \over r} {df \over d\epsilon} [({\bf \nabla}_\Omega \epsilon)
\times {\bf \nabla}] \}.
\label{pot}
\end{equation}
Here and in the following all derivatives of $f$ are meant to be calculated in
the limit $\epsilon=0$.

In this potential the first two terms correspond (apart from some different 
notation) to Eqs.~(3) and (5.1)
of Ref.~\cite{She}, and are usual spin--orbit potentials. The third term
refers to Eq.~(5.2) of Ref.~\cite{She}.

If one substitutes Eq.~(39) of Ref.~\cite{amo} (up to first order) into 
Eq.~(\ref{ls}), one 
obtains exactly the first two terms of Eq.~(\ref{pot}), namely:
\begin{equation}
V=V_{ls}{1 \over r}  [{df \over dr} +\epsilon
 {d \over dr} {df \over d\epsilon}] {\bf S \cdot L}. 
\label{pot1}
\end{equation}
However, in our approach we introduce also distortions to second order.
For simplicity, we introduce
distortions directly in the spin-orbit Eq.~(\ref{ls}) 
[instead of using Eq.~(\ref{base})]. In addition, we consider
a phenomenological spin--spin and also an orbit--orbit potential.
Note, in particular, that a spin--spin potential in our model was necessary
to have the correct separation between 
$J^\pi={3\over 2}^+$ and $J^\pi={5\over 2}^+$ resonances in the 
spectrum; an aspect extensively discussed by Tanifugi et al. 
in Ref.~\cite{res}.

Going to second order, namely assuming complete expansion as in
Eq.~(39) of Ref.~\cite{amo}, similar results are obtained. Eq.~(\ref{pot})
then becomes:
\begin{equation}
V_{th}=V_{ls} \{{1 \over r}  [{df \over dr} +\epsilon
 {d \over dr} {df \over d\epsilon}+\epsilon^2
 {d \over dr} {d^2f \over d\epsilon^2}] {\bf S \cdot L}-
{i \over r}  {\bf S}\cdot \{[{df \over d\epsilon}({\bf \nabla}_\Omega \epsilon)
+ {d^2f \over d\epsilon^2}({\bf \nabla}_\Omega \epsilon^2)]\times 
{\bf \nabla} \} \},
\label{pot2}
\end{equation}
and Eq.~(\ref{pot1}) assumes now the following form:
\begin{equation}
V=V_{ls} {1 \over r}  [{df \over dr} +\epsilon
 {d \over dr} {df \over d\epsilon}+\epsilon^2
 {d \over dr} {d^2f \over d\epsilon^2}] {\bf S \cdot L} .
\label{pot3}
\end{equation}
This corresponds exactly to the first part of Eq.~(\ref{pot2}), and is
the spin--orbit part of our interaction, which includes channel-coupling and
deformations.

The second part of Eq.(\ref{pot2}) is generally disregarded in calculations 
because it leads to 
a very complicated form and certainly it does not represent a conventional
spin-orbit (${\bf S}\cdot {\bf L}$-type) term. It can be shown that
it contains some type of spin-spin structure and note that
an effective spin-spin interaction is already included in our model potential.

Clearly, the form we use [see Eq.~(\ref{pot3})] is fully consistent with the 
${\bf S\cdot L}$ that comes from the Thomas term. That it can be considered
inconsistent or aberrant, as stated by Raynal, is simply false.
Note that we never claimed that the term in Eq.~(\ref{pot3}) represents
the fully complete spin structure of the deformed optical potential.
The model we use has indeed additional spin and orbital terms 
contributing to the spin-structure of the deformed optical potential. 

Now we come to the problem of the symmetrization of the potential.
Symmetrization is needed because our low-energy potential (which is
purely a real operator) is not symmetric when all the spin operators
(spin-spin, orbit-orbit and spin-orbit) are coupled with
the deformation operator. Thus symmetrization is needed to recover
unitarity in the S-matrix.

In absence of deformation the global interaction is given by Eq.~(36) 
of Ref.~\cite{amo}, with $f$ and $g$ given by Eqs.~(37) therein. When 
deformation is switched on, we may write in operatorial and schematic form:
\begin{equation}
V=f\  [V_0+V_{ll}{\bf L \cdot L}+V_{ss}{\bf S \cdot I}]+g\ V_{ls}
{\bf L \cdot S},
\label{uno}
\end{equation}
where now $f$ and $g$ are operators derived by introducing deformation into
the expressions $f(r,R)$ and $g(r,R)$ of Eq.~(37) of Ref.~\cite{amo}.


The expression of $\epsilon$ in Eq.~(38) of Ref.~\cite{amo} 
 implies\footnote{Note that in our case $L=2$}  that the operators $f$ and $g$
contain tensor terms of the type ${\bf Y}_2(\hat r) \cdot {\bf Y}_2(\hat s)$
and ${\bf Y}_4(\hat r) \cdot {\bf Y}_4(\hat s)$ ($\hat s$ standing for
the internal target coordinates) and  Eq.~(41) of Ref.~\cite{amo} gives the
detailed expression of $f$. In our representation, these terms
originate the channel couplings.

Since, in general, these tensor operators do not commute with the spin-- and
orbit--dependent operators, it is necessary to symmetrize Eq.~(\ref{uno}),
as follows:
\begin{equation}
V=V_0 \ f+{1 \over 2}V_{ll}[{\bf L \cdot L} \ f+f \ {\bf L \cdot L}]+
{1 \over 2}V_{ss}[{\bf S \cdot I} \ f+f \ {\bf S \cdot I}]+
{1 \over 2}V_{ls}[{\bf L \cdot S} \ g+g \ {\bf L \cdot S}].
\label{due}
\end{equation}
Finally, projection on states defined by Eq.~(35) of Ref.~\cite{amo}, and 
use of completeness properties, give rise to Eq.~(42) therein.
This expression is similar but more general
than the expression reported in Eq.~(7) of Ref.~\cite{She}.

It is important to note that we used this phenomenological model interaction
to test the sturmian algebraic method for narrow-resonance identification 
and to exhibit the relevance of the Pauli principle in the bound and resonance 
spectra.
For neutron-$^{12}$C scattering that interaction was sufficiently structured 
(but not exceedingly complicated)  to determine S-matrices from which
the {\em elastic} scattering data at very low energy was well reproduced.
Never did we claim use of a unique starting model interaction. Nonethelesss,
the terms involved (central, spin-orbit, spin-spin,
and orbit-orbit) are ones that typically appear in semi-relativistic 
reductions, $e.g.$ using, the Foldy-Wouthuysen transformation in 
atomic and molecular processes. We, and many others, have used them
in the nucleon-nucleus context in a purely phenomenological sense. 
These terms, however, do not saturate all the possible 
type of operators that can be constructed to describe semi-relativistic 
and many-body effects that can be possibly contained in this extremely 
complicated nucleon-nucleus optical potential. 
There is no claim in our work that the terms in our model calculation
represents ``the most general and complete form
of optical potential that can be constructed at a given order 
in the deformation
parameter''. That is, in our opinion, beyond present-day knowledge.

One should consider also that involved spin-spin and even tensor structures
naturally appear in the multinucleon interaction operators, and 
precisely in this particular symmetrized form,
if one combines pion-exchange dynamics with two-nucleon correlations
in a rigorous three-nucleon calculation. These specific forms
are known to have important effects for the vector analyzing powers
of nucleon-deuteron scattering at low energy\cite{OPE3nf}.

As a matter of fact, the interaction so defined is sufficiently structured and 
flexible with respect to parameter variation that a sound phenomenological 
analysis was possible, {\it provided that} 
the Pauli principle is taken into account 
and a powerful technique of resonance identification is used.
According to Raynal, the merit of our work is only to show that
some physicists are still using certain expressions for the deformed 
spin-orbit interaction, which in Ref.~\cite{She} is shown to give
slightly different results with respect to use of the ``complete'' 
Thomas term in DWBA calculations. But these effects are at higher energies 
and specifically for 
the analysing powers of the inelastic transitions. 

Raynal fails to realize 
that we use a model interaction 
which is much more rich in
structure (spin-spin, orbit-orbit, with addition of linear and quadratic 
deformation operators) with respect to what has been used in the references
he cites. Most importantly, our analysis is of the 
low-energy regime where only the elastic channel is 
accessible and the main issue
concerns the structure of the resonant spectrum as seen in the elastic process.
Here our analysis show that inconsistent results can be produced if
nonlocalities due to the Pauli principle are not taken into account, 
a problem that points directly to a serious flaw in the use of the 
ECIS formulation for such kind of problems.
This means simply that the relevance of the detailed structure of the deformed
spin-orbit potential, in this low-energy domain and 
for the kind of observable that we consider, is still 
{\em an open question}.\footnote{However, the results by Blair and Sherif
suggest that it is a question needing answer if one deals with 
inelastic spin observables and not with cross section}
Instead of producing
sterile polemics Raynal should try to overcome these 
methodological difficulties embedded in his ECIS calculational scheme.

\vskip 1 truecm
{\large \bf Conclusions}

We have replied to the criticism by Raynal in {\tt nucl-th/0312038}. 
Because the issue was poorly explained in that reference, 
we had first to make it intelligible. To summarise, the criticism is 
centered on the fact that we tested our calculation method 
within a model scheme  which contains a spin-orbit term Raynal
termed ``aberrant''. We showed in this reply that this is not the
case, since the spin-orbit expression that we use is fully consistent
with the ${\bf S\cdot L}$ term that comes from the full-fledged Thomas term.

In case of deformed nuclei, it is known that the ${\bf S\cdot L}$ term
does not exhaust all the possible spin structures of the
(coupled-channel) potential. In fact, to reproduce the data we 
phenomenologically included additional deformed (coupled-channel) spin-spin
and orbital-orbital terms, and carried over deformations to second order. 
Moreover, our coupled-channel potential is suited for very low-energy, 
and has to be hermitian. Because spin and/or orbital operators do not 
commute in general with the deformation operators, it has to be made 
hermitian by a symmetrization prescription. This puts a further constraint
in the spin structure, which has been duly taken into account.

To support the criticism, Raynal refers to old calculations.
Of those quoted calculations, only that by Sherif and Blair in 
Ref.~\cite{She} can be found with ease in the literature. 
These DWBA-type calculations refer to much 
higher energies (above 20 MeV), and specifically to the analysing powers 
for the {\em inelastic} process. Some differences were found
at forward angles for this specific observable, between results
obtained with the full
Thomas term  compared with those found using 
the symmetrized ${\bf L\cdot S}$ part of the Thomas term. 
However, Raynal seems to have missed that 
the very same reference warns that already at 18.6 MeV incident protons, 
this claim is not anymore fully consistent.

In conclusion, the criticism moved by Raynal against our analysis 
is {\em not} based on firm grounds because:

(i) our analysis is for the nucleon-carbon process at significantly 
lower energies where the main problem is specified by the structure 
of the resonance spectrum. Our computational method is particularly
designed for this kind of problem.

(ii) our analysis refers to elastic processes and observables, since
these are the only possible observables in nucleon scattering at 
such low energies. Instead, the effect of the full Thomas term
has been seen mainly at higher energies and for an inelastic scattering 
spin observable.

(iii) our model interaction is richer in its coupled-channel 
spin structure with respect to that analysed so far by the 
references quoted in {\tt nucl-th/0312038}. It contains additional
deformed spin-spin and orbit-orbit terms and all terms are carried 
to second order in the deformation parameter. For the specific
physical problem considered that structure has been proved
to be sufficiently rich to obtain a good reproduction of the resonant spectra,
of the background cross section, and of the elastic analysing power
below 4 MeV.

(iv) most importantly, our work has clearly shown
that phenomenological analyses such as those based on the ECIS
formulation are seriously flawed because they ignore
in the coupled-channel optical potential the effects due to the 
Pauli principle and, more specifically, of spurious states and resonances.
The coupling of the single-particle dynamics with collective-type
degrees of freedom of the target has been long known
to lead to an explicit violation of the exclusion principle.
This fact is clearly stated in old books~\cite{mahaux} 
as well as in widely used modern nuclear physics textbooks~\cite{Greiner}.
The combined use of the sturmian representation ($aka$ Weinberg states)
with the technique of orthogonalizing pseudo-potentials allows one 
to incorporate in the optical potential, formed by a collective
model of the target structure, the strongly nonlocal effects
of the projectile-bound nucleon indistinguishability, thus 
finally overcoming this notorious violation of the Pauli exclusion 
principle. Bound and super-narrow resonance spectra are strongly 
affected by this additional nonlocal term.
Without this additional nonlocal piece the corresponding spectra are largely
inconsistent, and this points to a serious limitation in the use of the ECIS 
method in phenomenological 
analyses for low-energy nucleon-nucleus scattering processes.
Given this problem, Raynal's polemics 
about the precise structure of the deformed spin-orbit potential in our 
low-energy neutron-$^{12}C$ test analysis suggests the curious situation 
of a person trying to call everybody's attention to a tiny mosquito
outside a closed window, while an elephant is sitting right in the middle
of his dining room.

\end{document}